\documentstyle[aas2pp4]{article}
\journalid{337}{15 January 1989}
\articleid{11}{14}

\begin{document}

\title{The Head-on Collision between Two Gas-Rich Galaxies: Neutral
Hydrogen Debris from the Centrally-smooth Ring Galaxy VII\,Zw\,466}

\author{P.~N.~Appleton\altaffilmark{1}, V.~Charmandaris\altaffilmark{1}
	and C.~Struck}
\authoremail{pnapplet@iastate.edu}

\authoraddr{Department of Physics and Astronomy, 
Iowa State University, Ames, IA 50011}
\affil{Department of Physics and Astronomy, Iowa State University, Ames, IA 50011}
\altaffiltext{1}{Visiting astronomer at NRAO. The National Radio
Astronomy Observatory is a facility of the National Science Foundation
operated under cooperative agreement by Associated Universities, Inc.}

\begin{center}
The Astrophysical Journal\\ submitted December 10 1995; accepted February
14, 1996\\ To appear on September 10, 1996
\end{center}

\begin{abstract}

We present VLA observations of the distribution and kinematics of the
H\,I gas in the classical ring galaxy VII\,Zw\,466 and its immediate
surroundings. The H\,I gas corresponding to the bright optical star
forming ring exhibits the typical profile of a rotating-expanding
ring. The systemic velocity of the galaxy is found to be 14,468$\pm$25
km\,s$^{-1}$. A formal fit to the HI kinematics in the ring is
consistent with both ring rotation and expansion (expansion velocity 32
km\,s$^{-1}$).  However, H\,I in the northeast quadrant of the ring is
severely disturbed, showing evidence of tidal interaction. In addition,
fingers of H\,I gas extend from the galaxy to the east in the general
direction of the two major companions.  We also detect a hydrogen plume
from the southern edge-on companion galaxy (G2) pointing towards the
ring galaxy. This, and other peculiarities associated with G2 suggest
that it is the intruder galaxy which recently collided with VII\,Zw\,466
and formed the ring.  Numerical hydrodynamic models are presented which
show that most of the observed features can be accounted for as a result
of the impact splash between two gas disks. The resultant debris is
stretched by ring wave motion in the bridge and later forms accretion
streams onto the two galaxies.

Finally, we detect H\,I emission from two previously unknown dwarf
galaxies located northeast and southeast of VII\,Zw\,466
respectively. This brings the total number of members of the
VII\,Zw\,466 group to five. Using the projected mass method, the upper
limit of the dynamical mass of the group was estimated to be $M_{o}$ =
$3.5 \times 10^{12}$ M$_{\sun}$, which implies that the mass to light
ratio of the group is (M/L)$_{group} \approx$ 70. This rather low value
of M/L, as compared with other loose groups, suggests the group may be
in a state of collapse at the present time. Plunging orbits would
naturally lead to an enhanced probability of head-on collisions and ring
galaxy formation.

\end{abstract}

%\twocolumn

\section{Introduction}

The idea that galaxies might be transformed as a result of interactions,
collisions and mergers was presented by Toomre \& Toomre (1972) to
explain the diversity of forms seen in many of the images of peculiar
galaxies (Arp 1966, Vorontsov-Velyaminov 1977). Observations since that
time have further strengthened the view that interactions play a
crucial role in galaxy evolution. In particular, as Toomre has
suggested, they may be responsible for the formation of elliptical
galaxies through major mergers and collisions. Observations of rich
clusters of galaxies at
redshifts of between 0.3 and 0.6 seem to show an increased fraction of
blue galaxies, with a significant fraction undergoing an interaction
(Butcher \& Oemler 1984; Lavery \& Henry 1988; Dressler {\em et al.}
1994). For those blue galaxies which are interacting, there is
increasing evidence from HST observations that many have shell-like or
ring-like morphologies (Lavery {\em et al.}  1995; Oemler
1995). Although these may not all be collisional systems, it is
interesting to study low-redshift galaxies which show similar
morphologies and ask whether these are simply analogs of processes
occurring at higher redshift. In this category we would include the shell
ellipticals (Malin \& Carter 1980), the galaxies with ripples (Schweizer
\& Seitzer 1988) and the collisional ring galaxies (Lynds \& Toomre 1976;
see Appleton \& Struck-Marcell 1996 for review ; hereafter AS96).

The collision of two gas-rich disk
systems of unequal mass ratio is probably one of the most challenging
problems in numerical modeling today (Barnes \& Hernquist 1992, Struck
1996a). The treatment of shocks in the disks of two inter-penetrating
galaxies is quite difficult to simulate and very few examples exist in
the literature. Struck (1996b) has recently attempted to model both the
hot and cold gaseous component during a head-on collision to produce the
Cartwheel ring galaxy and many complicated physical processes are at work
during and shortly after the penetration of the ``target'' galaxy by the
``intruder''. It is therefore of considerable interest to map the cool
and hot components of the gas in a real collision and make comparisons
with models. 

The ring galaxy VII\,Zw\,466 was discussed very early on in the debate about
the origin of these peculiar rings (Freeman \& De Vaucouleurs 1974;
Theys \& Spiegel 1976; Lynds \& Toomre 1976). An early photometric
study of VII\,Zw\,466 was made by Thompson \& Theys (1978) in which
the colors and knots within the ring were investigated. This galaxy has
also been the subject of a recent optical and infrared study (Marston
\& Appleton 1995; Appleton \& Marston 1995) and strong radial color
gradients were found within the ring.

Throughout this paper, we will adopt a Hubble constant of 80 km\,s$^{-1}$
Mpc$^{-1}$. We have determined the heliocentric optical velocity of the
galaxy to be 14,468 km\,s$^{-1}$ and therefore assume a distance to the
VII\,Zw\,466 group of 180 Mpc.

% In Section 3 we discuss the details of our observations

\section{Observations}

The observations were made on December 3 1994, using all 27 telescopes in the C
configuration of the VLA\footnote{ The Very Large Array is operated by
Associated Universities Inc. under cooperative agreement with the National
Science Foundation.} . The correlator was set in a two IF (Intermediate
Frequency) mode (AC~BD) with on-line hanning smoothing and 32 channels per IF.
For each IF we used a bandwidth of 3.125 MHz.  This provided a frequency
separation of 97.6 kHz per channel, which corresponds to 22.66 km\,s$^{-1}$ in
the rest-frame of the galaxy using the optical definition of redshift.  To
achieve a wider velocity coverage, IF1 was centered at 14,131 km\,s$^{-1}$ and
IF2 at 14,538 km\,s$^{-1}$. As a result, the velocity coverage of our
observations was 974 km\,s$^{-1}$.  A total of 4 hours 47 min was spent on
source. Flux and phase calibration was performed using the sources 3C286 and
01311+678 (B1950) respectively.

These data were first amplitude and phase calibrated and bad data due to
interference were flagged and ignored by the AIPS software. An image
cube was created from the UV data by giving more weight to those
baselines which sampled the UV plane more frequently (so-called natural
weighting). This provided a synthesized beam with a FWHM of $23.3\arcsec
\times 21.9\arcsec$ for IF1 and $21.6\arcsec \times 19.3\arcsec$ for IF2.

 The subtraction of the continuum emission in each line map was
performed using a standard interpolation procedure based on four
continuum maps free from H\,I at each end of the band. The resulting 
rms noise per channel was 0.27 mJy\,beam$^{-1}$.
The highest dynamic range in any channel map was 5:1 (at v=14,402.8
km/s) with a peak flux of 1.35 mJy/beam.  

 In order to determine the total H\,I distribution we used the
following technique.  Initially, we smoothed all channel maps to a
resolution $44.0\arcsec \times 44.0\arcsec$, twice that of the
synthesized beam. New maps were formed comparing, pixel for pixel, the
original full resolution maps to the smoothed ones. The pixel values of
original maps were copied to the new ones only if the signal-to-noise
ratio of the smoothed map at that point exceeded 2. The total H\,I
surface density map was produced by adding the new maps together.
The same technique was used to create the first and second moment maps
of the distribution. Using this procedure we effectively give more
weight to points associated with low surface brightness emission.

 Optical observations of a 16x16 arcmin field were obtained during photometric
conditions on April 2 1995, with the Fick Observatory 0.6 m telescope and CCD
system (Appleton, Kawaler \& Eitter 1993). A 500 sec exposure using an R--band
filter was taken. Calibration of the photometry was performed using data taken
from the star G10~50 (Landolt 1992)

\section{The VII\,Zw\,466 Group and its H\,I Distribution} 

 In Figures 1 and 2  we show the VII\,Zw\,466 group and its
environment. The ring has two nearby companions, labeled G1 and G2 
as well as a background galaxy B1 (Appleton \& Marston
1995). Galaxy B1 (first noted as background by R. Lynds; see Theys \&
Spiegel 1976) has a velocity of V$_{\sun}$ = 25,042 km\,s$^{-1}$,
determined from our own unpublished optical spectra.  Also marked in
Figure 1 are two new dwarf galaxies, G3 and G4, discovered during the
H\,I mapping of the group, presented in this paper.

In Figure 2 we show the integrated neutral hydrogen emission from
the VII\,Zw\,466 field superimposed on a B-band image of the group (from
Appleton \& Marston 1995). The emission from the ring galaxy is highly
disturbed. The brightest emission comes from the northern quadrant of
the ring and sweeps around in a clockwise direction along the western
quadrant of the ring. A filament of H\,I extends from the ring in a 
south-easterly direction,  pointing to a region between G1 and G2.
In addition, there is
a marked depression in the H\,I emission from the south-eastern quadrant
of the ring. It is in this quadrant that the optical emission in the
ring is rather faint. The H\,I distribution in VII\,Zw\,466 is similar
to that seen
by Higdon~(1993; 1996) in the Cartwheel ring galaxy, with scattered HI
extending in the direction of the three possible companions. As with
VII\,Zw\,466, the Cartwheel plumes are highly asymmetric, with all the
gas outside the galaxies being found on one side of the ring, highly
suggestive of a hydrodynamic splash resulting from the gas on gas
collision between the galaxies. 

H\,I emission is also detected from the edge-on companion G2. The H\,I
has a large peak centered on the south-eastern end of the major axis with
fainter emission extending along the major axis. A very faint filament
extends from G2 back towards the ring.

No emission down to a level of 0.68 mJy\,beam$^{-1}$ (2.5$\sigma$) was
detected from the peculiar elliptical galaxy G1. If we adopt a typical
velocity width of 300 km\,s$^{-1}$ for the assumed velocity of any gas
in the elliptical galaxy, this limit translates into a 2.5$\sigma$ upper
limit of $3.1 \times 10^{9}$ M$_{\sun}$ for the hydrogen mass of the
companion.

We also detected gas from two new group members G3 and G4 located
3.2\arcmin~ and 4.2\arcmin~to the southwest and northwest of
VII\,Zw\,466, respectively. At the distance of the group these correspond
to projected linear separations of 166 kpc and 220 kpc from the ring.

\section{Integrated Properties}

In Table~1 we present the H\,I and optical properties of VII\,Zw\,466, its
two major companions (G1 and G2), and the two newly discovered dwarf
galaxies, G3 and G4. The R--band luminosites, L$_{R}$, of G1 and G2
are from Appleton \& Marston (1995) while the optical properties of G3
and G4 were determined from the Fick CCD images. There is excellent
agreement between the photometry of the galaxies in common between the
previously published KPNO magnitudes and the Fick magnitudes.

The velocity profiles of the neutral hydrogen distribution of
VII\,Zw\,466, G2 and two dwarfs are presented in the
Figures~3--6. The H\,I profile of VII\,Zw\,466 presents
the typical two-horned profile of a rotating disk. We estimate the
systemic velocity of the ring to be 14,468 $\pm$ 25 km\,sec$^{-1}$, in
close agreement with the optically determined velocity of 14,490
km\,sec$^{-1}$ (Theys \& Spiegel 1976). However, the high velocity horn
is not completely symmetrical.  Emission seen at velocities higher than
14,520 km\,sec$^{-1}$ is associated with the gas plumes outside of the
visible galaxy.

 The H\,I profile of G2 is single-peaked with a $\Delta V_{1/2}$ = 84
km\,sec$^{-1}$. The galaxy appears to be edge-on, so for all calculations we
used an inclination $i$=90\arcdeg.  The two dwarfs, though, have
more peculiar H\,I profiles.  The velocity profile of G3 exhibits a
low-level asymmetry around
14100 km\,sec$^{-1}$ which may not be real. For this reason, we did
not attempt to calculate $\Delta V_{1/5}$ for this galaxy. Finally, G4, the
least massive galaxy in the group, has a very narrow H\,I profile with
$\Delta V_{1/2}$ = 25.3~ km\,sec$^{-1}$. For both G3 and G4 we assumed
an inclination of $i$=45\arcdeg, since the disks of the galaxies were
only marginally resolved in the optical images of the group. 

 We also include in Table~1 a rough estimate the dynamical mass
M$_{d}$ of each group member. The values of M$_{d}$ corresponding to
the galaxies G2, G3 and G4 were calculated using the formula:

$$
{M_{d}} = 
{\frac{(\frac{1}{2} \Delta V_{\frac{1}{2}} cosec(i))^{2}R_{HI}}{G}}
$$
which assumes that the gas is in bound circular orbits. Values of the H\,I
radius were obtained by inspection of the run of emission centroid
with radius for each galaxy and in all cases the H\,I disks were
resolved and a radius determined. 

 The estimate of the dynamical mass of VII\,Zw\,466 was produced by the
fit to the rotation of the H\,I ring discussed in Section 6.  For
completeness, the dynamical mass of G1 was simply estimated from its
observed luminosity, using an R--band mass to light ratio of M/L=15,
typical of elliptical galaxies (Faber \& Jackson 1976).  Finally, we
present the mass to light ratio M$_{d}$/L$_{R}$ of G2, G3, G4, and
VII\,Zw\,466. We note the surprising small values of M$_{d}$/L$_{R}$ as
well as M$_{HI}$/L$_{R}$ for the spiral galaxy G2. We believe that this
may be related to the tidal stripping of its mass. We will elaborate more on
that in Sections 6 and 8.

\section{The Dynamical Mass of the VII\,Zw\,466 Group}

 Our observations indicate that there are now four galaxies G1--G4 which
form a loose group around VII\,Zw\,466. Since we know the recession
velocities of those galaxies and their angular separations, we can use a
dynamical method to get an independent estimate on the total mass of the
group and subsequently, its dark matter content. This is the first time
that the large-scale matter content of a group containing a ring
galaxy has been estimated.

The primary method used is the {\bf projected mass estimator} proposed
by Bahcall \& Tremaine (1981). However, for comparison we also estimated
the mass of the group using a mass weighted virial theorem
approach. Both give similar results.

The projected mass method is based on the idea that, for a dynamically
bound system with one massive object and several smaller companions, one
can examine two extreme cases for the possible orbital types of the group
members. The first case assumes an isotropic distribution in the velocities
of the members which implies that the average value for the eccentricity
of the orbits is $<e^{2}>= 1/2$. The other extreme is that all orbits of
the group members are radial implying that $<e^{2}>= 1$. These lead us
to define two specialized estimators of the dynamical mass:

$$
M_{I} = \frac{16}{\pi G N} \sum_{i=1}^N v^{2}_{zi} R_{i}
$$

and

$$
M_{R} = \frac{32}{\pi G N} \sum_{i=1}^N v^{2}_{zi} R_{i}
$$

where $v_{zi}$ is the difference in the recessional velocity between the
group member $i$ and the central massive object and $R_{i}$ is its
projected distance from the central object (see Bahcall \& Tremaine
1981, for details).

When one has no specific information on the distribution of the
eccentricities of the group one can use a dynamical mass estimator which
is the average of the two extremes, that is:

$$
M_{o} = \frac{1}{2}(M_{I}+M_{R}) = 
	\frac{24}{\pi G N} \sum_{i=1}^N v^{2}_{zi} R_{i}
$$

In the VII\,Zw\,466 group the most massive object is the elliptical galaxy G1.
However, since its velocity is poorly determined (no H\,I detection) and
its mass is inferred from its luminosity and an assumed mass to light
ratio, we decided to to use the barycentric velocity and position of the
group as the point of reference, rather than G1.  Using the figures
presented in Table~1, we find that the barycenter is located at
$\alpha_{cm}$=12\fh 32\fm 11.2\fs~and $\delta_{cm}$=+66\arcdeg 23\arcmin
58.51\arcsec~ (J2000) and its systemic velocity is $v_{cm}$=14,143
km\,s$^{-1}$. The new position for the center lies only 1 galactic
radius away from the elliptical (the elliptical still dominates). Our
results for the mass of the group are presented in Table~2.

 Although the use of the barycenter rather than a single galaxy as the center
may seem to contradict the assumption that the group is dominated by a
single central massive object, we feel that this is a compromise
position, given the poorly determined optical velocity of G1. If the
group is dominated by dark matter, then the barycenter of the group
might be more appropriate anyway. We note for comparison that had we
used G1 as the center, our estimate of the dynamical mass would only be
greater by a factor of 1.83 than the result presented in Table~2.

 The estimate of the total dynamical mass of the group is
$M_{o}$\,=\,$3.5 \times\,10^{12}$ M$_{\sun}$. The sum of the dynamical
masses of each individual ($ith$) galaxy from Table~1 is:

$$
M_{T} = \sum_{i=1}^{5} M_{di} = 3.55 \times 10^{11} M_{\sun}
$$

which implies that the amount of dark matter distributed outside the 
members of the group is: 

$$
\frac{M_{o}}{M_{T}} \approx 9.9
$$

Table~1 also shows that the total luminosity of the group is
L$_{RT}$=5.02 $\times$ 10$^{10}$ L$_{\sun}$, implying that the total
mass to light ratio for the group is (M/L)$_{group} \approx$ 70.

For completeness we also calculated the mass of the group using a method
based on the mass-weighted virial theorem. In this approach the mass of
the group is given by the formula:

$$
M_{VT} = \frac{3 \pi}{2 G} 
	 \frac{\displaystyle \sum_{i=1}^5 M_{di} v_{zi}^2}
	      {\displaystyle \sum_{i=1}^5 \sum_{j=1, j>i}^5 
			\frac{ M_{di} M_{dj} }
			     {r_{ij}} 
	      }
	 \times \sum_{i=1}^5 M_{di}
$$

where $M_{di}$ is again the dynamical mass of a group member $i$ and
$r_{ij}$ is the the projected distance between the members $i$ and $j$.
This results to a group mass $M_{VT}$ = 6.2 10$^{12}$ M$_{\sun}$, which
is only 1.78 times larger than the projected mass method estimate.  

We believe that the projected mass method estimate is more accurate than
the one resulting from the virial theorem technique. Bahcall \& Tremaine
(1981) have shown through numerical simulations, that the virial theorem
technique may be biased and inefficient in particular for systems with
small number of members.  In addition, the projected mass method has the
added advantage that it explicitly takes into account the possibility
that some or all of the orbits in the group are radial. The latter reason
is very relevant in this study. Unlike most group studies, we know that
in the VII\,Zw\,466 group at least two of the galaxies (perhaps G2 and
VII\,Zw\,466) must have collided radially in order to produce the ring.

The projected mass method and the virial mass method are valid only if
the group is gravitationally bound. If some of the galaxies are
leaving the group, or are chance projections of foreground or
background galaxies, or the group is not stable, then the resulting
dynamical mass and group crossing times would be incorrect.

Other studies of loose groups of galaxies (e.g. Ramella, Geller \&
Huchra 1989) show that they have a typical (M/L) $\approx$ 150-300,
while in compact groups (M/L)$\approx$ 50 (Hickson {\em et al.}
1992). In some scenarios, it is believed that compact groups are
believed to be continously forming at the center of rich loose groups
and are in a state of collapse. The fact that the mass to light ratio of
the VII\,Zw\,466 group is similar to that of compact groups and also the
presence of a least one observed collisional ring galaxy pair in the
group, suggests that, like the compact groups, VII\,Zw\,466 group may
also be collapsing.

\section{The H\,I Kinematics of the Ring}

 In order to understand the kinematical behavior of the H\,I gas in
the ring of VII\,Zw\,466, we used a model which assumes that the ring
may be both rotating and expanding. This model has been used
successfully in other ring galaxies, such as the Cartwheel (Fosbury
and Hawarden 1977; Higdon 1993).  Using the optical B--band image of
the group and assuming an intrinsically circular ring\footnote{We
caution that we have assumed that the ring is an intrinsically
circular structure and that its elliptical shape is entirely a result
of an inclined viewing angle. We acknowledge that in models of
off-center ring formation, in the early expansion stages,
the ring is somewhat non-circular. This may lead to an
overestimation of the inclination of the ring and would require a more
complicated treatement of the kinematic data.}, we estimated the
inclination of the ring to be $i=$ 37\arcdeg~and the position angle of
the major axis to be -5\arcdeg~(N through E).  Then, we measured the
velocities of the H\,I gas along azimuth of the ring and performed a
three parameter fit using the following function:

\begin{equation}
v = a_{0} + a_{1}sin(\phi+\phi_{0})
\end{equation}

where $a_{0}, a_{1}$ and $\phi_{0}$ were the free parameters. A rotating
and expanding ring would exhibit a simple sinusoidal shape, with a phase
offset from the major axis that is related to the amplitude of the
expansion. The formal fit for the ring resulted in a $\chi^{2}$=5.61.
However, the points of the northeast quadrant of the ring exhibit a
peculiarity related to one of the H\,I plumes, and the errors associated
with them are larger. If we exclude those points, (which is reasonable
since we are only trying to model the ring) the fit improves
considerably, giving a $\chi^{2}$=2.77 for $a_{0}$=14,500 km\,s$^{-1}$,
$a_{1}$=83~km\,s$^{-1}$, and $\phi_{0}$=274\arcdeg. This fit is
presented in Figure~7. The azimuthal component of the
velocity at the ring radius and the radial expansion velocity, corrected
for inclination are found to be $v_{max}$=137~km\,s$^{-1}$ and
$v_{exp}$=32~km\,s$^{-1}$.  We note that this expansion velocity is
significantly larger than the 8~km\,s$^{-1}$ obtained optically by Jeske
(1986). We believe that the combination of the two-dimensional nature of
our fit and the high velocity resolution of our observations makes our
measurement more reliable.

In Figure~8 we also present the channel maps of the H\,I
observations. In the rest frame of the galaxy, the channel separation
is 22.66 km\,s$^{-1}$. H\,I emission from the VII\,Zw\,466 is detected
for the first time around 14,561 km\,s$^{-1}$ and originates from the
southern side of the galaxy. As we move to lower velocities,
the emission is typical of a rotating ring. We observe that the
emission splits in two centroids located to the east and west of the
optical ring. However, the eastern centroid does not overlap spatially
with the optical picture of the ring. It resides at the outside of the
optical ring structure and at 14,470 km\,s$^{-1}$ is deformed into a
small plume (hereafter ``Plume~A'') pointing to the southeast. The
amount of H\,I gas associated with this plume is M$_{A}$=1.6 $\times$
10$^{8}$ M$_{\sun}$. The emission from the western centroid does
coincide with the optical ring. It is also considerably stronger,
indicating large quantities of gas. Strong emission from the west side
of the ring was also detected in the radio continuum observations of
the galaxy (Ghigo \& Appleton, in preparation).

As we proceed to channel maps with velocities below 14,470 km\,s$^{-1}$,
the behavior of the gas is very disturbed. Firstly, we notice that at
14,448 km\,s$^{-1}$ there is practically no H\,I gas associated with the
ring. This is very peculiar since there is a strong emission in the
preceding and following channel maps. Also, there is a filament
(hereafter ``Plume B'') of H\,I gas present in three consecutive channel
maps (14,402 km\,s$^{-1}$ -- 14,357 km\,s$^{-1}$). This is the strongest
emission that we observe. The filament initially points to the northeast
of the ring and as we progress to lower velocities, it points to the
east-southeast of the ring in the general direction of the companions G1
and G2. We calculated that the H\,I gas associated with plume B is
M$_{B}$=4.2 $\times$ 10$^{8}$ M$_{\sun}$.
 
At this point, it also useful to discuss the H\,I emission from the
companion G2. The emission from G2 is located in the channel maps
between 14,516--14,334 km\,s$^{-1}$. However, we notice H\,I gas
at a velocity of 14,584 km\,s$^{-1}$ and a lack of H\,I gas at 14,425
km\,s$^{-1}$. Moreover, the emission centroid from G2 in the channel maps
after 14,425 km\,s$^{-1}$ is rather elongated pointing clearly towards
the ring galaxy. Table 1 shows that the value of M$_{HI}$/L$_{R}$ for
G2 is also unusually low for a late-type spiral, suggesting that it has
been tidally stripped. (We also note that the total mass to light ratio
of G2 is extremely low, implying that the dark matter halo of G2 has
also been stripped). 

The dynamical behavior of the H\,I gas is consistent with G2 being the
galaxy that disturbed the progenitor H\,I disk of VII\,Zw\,466 and
caused the formation of an expanding star forming ring. Numerical
simulations of the gas dynamics of a nearly head-on collision between a
large gaseous disk galaxy and a smaller companion disk galaxy (Struck
1996a;1996b; see also Section~8) show that large quantities of gas are
stripped away from the disk of the target galaxy by the companion during
the collision.  Most of the ejected material either falls back to the
large disturbed disk or it spirals towards the smaller companion which
eventually accretes it. We will argue that the asymmetric nature of the
H\,I distribution around G2 is strongly suggestive of the onset of
accretion from one such tidal stream.

Plumes A and B and emission which extends from the southern end of
VIIZw466 (see Figure 2) probably form part of a two-horned H\,I bridge
between VII\,Zw\,466
to G2. This rather special morphology is predicted by the models and
strongly suggests that the plumes are splattered debris from the
collision. The total amount of H\,I gas in plumes A and B is
M$_{A}$+M$_{B}$ =5.8 $\times$ 10$^{8}$ M$_{\sun}$, which is
approximately 16\% of the mass of the H\,I disk of VII\,Zw\,466. A
more detailed discussion of the formation of these structures is given
in the next section.

To further estimate the energy needed to generate the two H\,I plumes,
we performed an order of magnitude calculation of the kinetic energy of
the gas in the plumes. We used the H\,I masses given above and the
relative velocities of plumes A and B $\Delta v_{A}$ = 14 km\,s$^{-1}$ and
$\Delta v_{B}$ = 88 km\,$s^{-1}$ with respect to VII\,Zw\,466. The kinetic
energy associated with their gas was found to be KE$_{A}$= 3.1 $\times
10^{53}$ ergs and KE$_{A}$= 3.2 $\times 10^{55}$ ergs respectively.
Although projection effects are not included, this implies the energy
required to displace the plumes from VII\,Zw\,466 is of the order of
10$^{55}$ ergs. This is approximately 10\% the gravitational potential
energy of the VII\,Zw\,466 -- G2 pair, indicating that the collisional
scenario is plausible.
 
\section{The Possible Role of the Elliptical Companion G1}

In our discussion so far we presented evidence in
favor of the edge-on disk-like companion G2 as the most likely intruder
galaxy. Can we completely rule out the possibility that G1, the
brighter elliptical galaxy, is the galaxy that plunged through the
center of the progenitor disk of VII\,Zw\,466? After all, an
inspection of Figure 1 and 2 shows that the elliptical system lies
closer to the minor axis of the ring, is slightly closer to the ring
(2.5 rather than 3 ring diameters away) and has been shown to have
peculiar optical isophotes (Appleton and Marston 1995). In addition,
its optical velocity (though poorly known) differs from the ring
velocity
by over 350
km\,s$^{-1}$, suggesting a high
impact velocity with respect to the ring.  

Although we cannot completely rule out the possibility that the
elliptical galaxy is the intruder galaxy, we believe the evidence at
present is against it. Firstly, all our simulations to date lead to
the formation of an accretion stream, which is a bridge between the
two galaxies involved in the collision. In the only similar
21cm HI study so far published, the same behavior is found for
the Cartwheel galaxy, where a long HI streamer connects the Cartwheel
to its most distant disk companion (Higdon 1996).  The fact that the
elliptical system is not connected to the ring, nor shows any evidence
for HI down to a low column density (see Table 1) seems to rule out
accretion of cool gas onto galaxy G1. A careful analysis of the color
map of the VIIZw466 system (see Appleton and Struck-Marcell 1996)
seems to rule out any evidence for a dusty disk inside G1, similar to
those seen in other ellipticals in which accretion is believed to be
taking place (e.g. Sparks et al. 1985). Is it possible that any
accreting gas might have become ionized? If so, such emission must be
weak, since imaging in H$\alpha$ by Marston and Appleton (1995) failed
to show any evidence for ionized hydrogen. In addition, unpublished
radio continuum maps of the system by Ghigo and Appleton (in
preparation) show that both VII\,ZW\,466 and G2 are strong radio
emitting objects at a wavelength of 6cm, but that no emission is
observed from the elliptical galaxy G1.
Similarly, Thompson and Theys (1978) found comparably blue colors in
VII\,Zw\,466 and G2, but not in G1. These points again suggest that G1 is
a rather inactive object and probably has not suffered accretion
recently. 

Perhaps the strongest arguments in favor of G1 being the intruder
relates to the position of the galaxy relative to the minor axis of
the ring and its high relative velocity compared with the
ring\footnote{Indeed, although Theys and Thompson (1978) seemed to
favor G2 as the intruder based on color evidence, they could not rebut
this argument in a convincing way (They appealed to three-body
effects).}. The galaxy G1 is only a little closer to the ring than G2,
and so from this point of view neither is favored. N-body simulations
by Huang and Stewart (1988) and Appleton and James (1990) show that
very respectable rings can be generated in collisions of quite low
inclinations between the intruder and the target disk, reducing the
need to always find the intruder close to the minor axis of the ring
in all cases. Even the high relative velocity is not a strong argument
for identifying G1 as the intruder. Again N-body simulations of ring
galaxies (e. g. R. A. James, private communication) show that
moderately unbound collisions are turned into bound systems via
dynamical friction and that it is likely that the relative velocity of
the intruder would be significantly reduced after it has travelled a
few ring diameters from the target (as observed). In conclusion, we
believe that the  HI data provide strong evidence that the disk galaxy
G2, rather than the elliptical G1, is the likely intruder in this
system.

\section{Models}

The H\,I distribution of Figure~2 suggests that we are witnessing the messy
aftermath of a collisional splash between two galaxies.
We present here a numerical simulation which
supports this picture, and provides further details about the
dynamics. The model presented here is not a detailed model of this
particular system, but should be viewed as a working example of how
the collisional scenario (disk plus disk collision) can produce much of
the observed structures. Until higher resolution HI
observations are available, a more detailed model which tries to fit
all aspects of the system perfectly is not yet justified. The model
presented here is part of larger program of simulations between two
gaseous galaxy disks.  Some preliminary results have been presented in
AS96 and Struck (1996a), and a journal paper on these simulations is
in preparation (Struck 1996b).

The simulation code is described in the above references, and an earlier
version is discussed in Struck-Marcell \& Higdon (1993). In this code,
the large-scale collision dynamics are modeled within the restricted
three-body approximation, which is adequate for the early transient
stages of galaxy collisions (e.g. Gerber \& Lamb 1994).  In these models
each galaxy consists of a rigid halo and a gas disk.  The potential of
the primary galaxy is such as to give a rising rotation curve, which
becomes nearly flat at large radii, and extends to several disk
diameters.  A simple softened, point-mass form is used for the companion
potential.  The gas dynamics is computed with a smooth particle
hydrodynamics (SPH) algorithm with a spline kernel.  In addition to the
force from the large-scale potential, a local self-gravity is computed
between neighboring gas particles within the disks, which allows the
formation of clumps or clouds.  Shear dominates over local gravity on
scales larger than that on which the latter is calculated.

Simple models for heating and cooling are also included in this
simulation.  However, isothermal comparison runs show that the
large-scale morphology is little effected by these terms (see Struck
1996b for details).  The most important result of including these terms,
for present purposes, is that the simulations show a prompt, adiabatic
cooling of bridge material.

The two disks were initialized in centrifugal balance, with the
companion set several primary-disk diameters from the target, and
with a slight sideways velocity to allow a slightly off-center impact.
The x-y plane is the initial midplane of the primary disk, while the
companion midplane lies on a translated y-z plane.  The companion
orbits in the x-z plane.  There is sufficient time before
impact for the heating/cooling terms of both disks to reach a
quasi-equilibrium.  

At the time of impact, extremely strong shocks form.  A good deal of primary
material is splashed out, so that the bridge ultimately contains a
fairly even mix from both progenitors.  Similarly, much gas from the
companion is left behind.  Although this simulation was not specifically
produced to model VII\,Zw\,466, the model and timestep shown
in Figure~9 were chosen as a reasonably good match. In a
comparison model using a gasless companion but otherwise identical
parameters to the current model show that no gas bridge was formed and
little gas debris was ejected from the gas disk.
Thus, we see that collisional hydrodynamics (especially a gas on gas
collision) seems to plays a key role in the
formation of the double-horned HI bridge, in contrast to bridges formed
in prograde, planar interactions (e.g. Toomre \& Toomre 1972).
VII\,Zw\,466 is apparently an interesting example of such a highly dissipative
gas-dynamical collision.

Figure~9a,b gives two orthogonal views of the system some time
after impact, when the companion has moved a good distance away, with a
prominent bridge stretching between it and the primary.  We have not
attempted to find an optimal viewing angle for the model, though it
appears that one about half-way between the two orthogonal views would
be quite good.  In any case the morphological similarities between
Figures 2 and 9 are quite strong.  First of all, both have a similar
double-horned bimodal appearance, or alternately, a central hole in the
bridge material.  Animated views of a sequence of model timesteps
provide a simple explanation.  The ring galaxy itself is produced by first an
inward gravitational impulse followed by a centrifugal rebound.  The
same kinematics occurs in the bridge material, with the addition of
stretching between the galaxies.  Thus, in this picture the bridge is
essentially a ring stretched the vertical direction, or a cylindrical
annulus.

Figure~9a,b also shows the pattern of infall onto the two galaxies.
In the case of the primary disk there is significant infall out of both
streams, and the accretion tends to be concentrated in the central
regions.  In the case of the companion the (highly supersonic) accretion
out of one stream is much stronger than the other, though the details
are quite time-dependent.  In both cases the accretion streams swirl
around the nucleus and then merge into the ambient rotational flow.
Figure~9a shows that gas swirls out a relatively large distance
on the side opposite the main stream, which may account for the HI
asymmetry of G2 in Figure~2.  The models indicate that the accreting
material is shocked after circling nearly all the way around the
companion, and there is much hot and warm gas concentrated on the main
stream side.  The companion is a radio continuum source, and the models
predict that this emission would be offset from the galaxy center if
arises from such shocks. This is in fact observed (Ghigo and Appleton,
in preparation). In the model, the companion disk is highly disrupted in
the collision, and largely reforms in the accretion process.
Although we have no direct evidence for accretion onto VII\,Zw\,466
itself, we note that the large circular loop seen at optical wavelengths
in the north-west quadrant of the ring might be the result of bridge
material falling back onto the disk. Such material might induce star
formation in a ring, which is consistent with H$\alpha$ observations of
the loop (Marston \& Appleton 1995).

In an offset collision, the primary disk receives an azimuthal swing
impulse as well as the radial one.  This leads to asymmetric and
offset spatial distributions as seen in both Figure~2~and ~9b.
Figure~9b also shows a developing oval ring rather like that in
VII\,Zw\,466.  This is deceptive, however, since it is the second ring
in the model, with the first ring moving out of the disk.  At present
there is no independent evidence to suggest that the ring in
VII\,Zw\,466 is not the first.  Thus, it appears that the impact
velocity in VII\,Zw\,466 was high relative to orbital or epicyclic
speeds in the primary disk, because the companion has moved several
ring diameters.  The model shows a similar morphology with the first
ring, but the bridge has hardly developed at that time. With the aid
of kinematical constraints from higher-resolution observations at
21cm, future models for this system should be able to resolve this
timing problem, and provide more specific predictions.

\section{Conclusions}

 In the previous sections we presented our H\,I observations of the
VII\,Zw\,466 group. Though only at moderate resolution due to the large
redshift (z=0.048) of the group, we were able to draw the following
conclusions~:

\begin{itemize}

\item The H\,I emission from VII\,Zw\,466 is  consistent with that of a
rotating and expanding ring. A simple kinematical model fits the ring
sufficiently well and indicates a rotation curve with $v_{max}$ = 137
km\,s$^{-1}$ at the radius of the ring, and an expansion velocity
$v_{exp}$=32~km\,s$^{-1}$. However, one quadrant of the ring has
anomalous velocities and is associated with disturbed H\,I outside the ring.

\item  The northern part of the ring shows evidence of the tidal
interaction. Plumes of H\,I gas (5.8 $\times$ 10$^{8}$ M$_{\sun}$, or
16\% of the total H\,I disk) lie outside the ring and point in the
general direction of the two nearby group members. The dynamics of this
gas is consistent with a collisional origin for the ring.

\item Of the two nearby companions, only the edge-on spiral, G2, contains
neutral hydrogen above our detection limits. Its H\,I properties are abnormal.
Not only does G2 exhibit a plume which points towards the VII\,Zw\,466, but it
also has very low values of  M$_{HI}$/L$_{R}$ and M$_{d}$/L$_{R}$. This strongly
suggests that G2 is the intruder galaxy that collided head-on with VII\,Zw\,466
and created the ring (Lynds \& Toomre 1976). We conclude that the elliptical
galaxy G1 has played little role in the ring formation, despite its unusual
``boxy'' optical isophotes (Appleton \& Marston 1995).

\item  Models suggest that the plumes represent a mixture of gas ejected from
the progenitor of the ring galaxy and material stripped from the suspected
intruder galaxy. The hydrodynamical models provide a good match to the plume
morphology and suggest that accretion onto both galaxies is already underway.
The models also suggest the possibility that the companion disk was strongly
disturbed in the collision, and that the observed asymmetries are the result of
partial reformation of the (G2) disk in the accretion process.

\item We identified two new members of the extended VII\,Zw\,466 group. These
galaxies are dwarfs and are detected for the first time in H\,I.

\item An upper limit to the dynamical mass of the group was estimated
using two different techniques and was found to be in the range $M_{o}$
= $3.5$ to $6.2 \times 10^{12}$ M$_{\sun}$. Adopting the lower end of
the range as being more reliable, this implies that the dark matter not
associated with the group members is 10 times greater than the sum of
the dynamical masses of all group members. The resulting mass to light
ratio for the group is (M/L)$_{group} \approx 70$.

\item The estimated M/L for the group is roughly a factor of three
times lower than the median M/L for loose groups, but is similar to
that estimated for compact groups of galaxies. As with compact
(Hickson) groups, the low M/L ratio may imply that the group is not in
equilibrium, but may be collapsing. Such a scenario would  naturally give
rise to an increased incidence of deeply penetrating orbits and
might increase the probability of a head-on collision leading to the
formation of a ring galaxy. 

\end{itemize}

 We note that, if the numerical models are correct, the gas plumes seen
in these H\,I observations may represent only a part of the debris
formed as the intruder collided with the disk of VII\,Zw\,466. If the
collision was quite recent (as implied by the fact that only one ring is
seen in VII\,Zw\,466 -- see AS96 for review), then it is possible that much
of the shocked gas accreting onto the two galaxies is still strongly
ionized. Deep H$\alpha$ observations or high sensitivity UV or X-ray
observations may be fruitful.

\acknowledgments

The authors would like to thank R.J.~Lavery (Iowa State University) for
stimulating discussions on general galaxy interactions, and J. J. Eitter
for instrumentational support at the Fick Observatory. The authors are
grateful to an anonymous referee for helpful comments on the
manuscript. We are also
grateful to E.~Brinks (NRAO, Socorro) for useful suggestions during the
VLA data reduction process. This work is funded under NSF grant
AST-9319596.

%
% The references
%

%% 
%% Table 1 
%%

\begin{deluxetable}{lccccc}
\tablewidth{0pc}
\tablecaption{Properties of the VII\,Zw466 Group}
\label{7zwtbl}
\startdata\nl
\tableline \nl

& VIIZw466 & Elliptical (G1) & Spiral (G2) & Dwarf-1 (G3) & Dwarf-2 (G4)\nl
\tableline \nl
$\alpha$(J2000) & 12\fh 32\fm 05.27\fs &
                12\fh 32\fm 13.30\fs &
                12\fh 32\fm 11.95\fs &
                12\fh 31\fm 37.99\fs &
                12\fh 31\fm 35.11\fs \nl
$\delta$(J2000) & +66\arcdeg 24\arcmin 13.65\arcsec &
                +66\arcdeg 23\arcmin 59.80\arcsec & 
                +66\arcdeg 23\arcmin 19.92\arcsec &
                +66\arcdeg 22\arcmin 35.56\arcsec & 
                +66\arcdeg 26\arcmin 55.26\arcsec \nl
%Distance (D) & 180 Mpc & 180 Mpc & 180 Mpc & 157 Mpc &210 Mpc  \nl

V$_{HI}$ (km\,s$^{-1}$) &  
		14,468 $\pm$ 25& [14,100]\tablenotemark{a}
\tablenotetext{a}{ optical velocity from Theys \& Spiegel (1976)}
		& 14,457 $\pm$ 25 & 14,046 $\pm$ 25 & 14,223 $\pm$ 25\nl

$\Delta V_{1/2}$ (km\,s$^{-1}$) & 
		202  &\nodata & 84 & 113 &25  \nl 

$\Delta V_{1/5}$ (km\,s$^{-1}$) & 
		256  & \nodata & 119  &  \nodata & 40  \nl 

R$_{H\,I}$ (arcsec) & 11 &\nodata & 6.7 & 9.3 & 4.6\nl

$\int S(v)dv$\tablenotemark{b}
\tablenotetext{b}{where $\int S(v)dv$ is the integrated 
                flux density in Jy\,km\,s$^{-1}$ } 
	& 0.541 & 
        $\leq$ 0.408\tablenotemark{c} 
\tablenotetext{c}{assumes a hypothetical velocity width of 300 km\,s$^{-1}$ and 
	a solid angle equal to two VLA beams} & 
        0.155 & 
        0.383 &
        0.092 \nl

M$_{HI}$/M$_{\sun}$\tablenotemark{d}
\tablenotetext{d}{
                where M$_{HI}$/M$_{\sun}$ = D$^{2} \times 
                2.356 \times$ 10$^{5}\int$ S(v)dv  and D = 180 Mpc
        } & $4.1\times10^{9}$ & 		% $4.136\times10^{9}$ & 
        $\leq 3.1\times10^{9}$$^{c}$ &  
        $1.2\times10^{9}$ &  			% $1.185\times10^{9}$ & 
        $2.9\times10^{9}$ & 			% $2.928\times10^{9}$ & 
        $7.0\times10^{8}$  \nl 			% $7.026\times10^{8}$  

R--band mag.\tablenotemark{e}
\tablenotetext{e}{from Appleton \& Marston (1995)}
		& 14.94 &14.64 & 15.62 & 16.18 $\pm$ 0.05 & 17.67 $\pm$ 0.10\nl

M$_{d}$\tablenotemark{f}
\tablenotetext{f}{M$_{d}$ = $\frac{(0.5 \Delta V_{\frac{1}{2}} cosec(i))^{2}R_{HI}}{G}$
	is an estimate of the total dynamical mass of the galaxy.}
	 	(M$_{\sun}$) &
		4.10 $\times 10^{10}$ & 
		3.00 $\times 10^{11}$ & 
		2.31 $\times 10^{9}$ & 
		1.16 $\times 10^{10}$\tablenotemark{g}& %5.8 x sqrt(2)
		1.80 $\times 10^{8}$\tablenotemark{g} \nl %1.4 x sqrt(2)

\tablenotetext{g}{since the dwarf was unresolved from the B--band image
we assumed an inclination $i=$ 45\arcdeg}

L$_{R}$ (L$_{\sun}$)     & 1.6$\times 10^{10}$ 	%& 1.58$\times 10^{10}$ 
		& 2.0 $\times 10^{10}$ 
		& 8.2 $\times 10^{9}$
	 	& 4.9 $\times 10^{9}$
		& 1.3 $\times 10^{9}$\nl

M$_{HI}$/L$_{R}$ (M$_{\sun}/$\L$_{\sun}$) & 
		0.26 & $\leq$ 0.16 & 0.14 & 0.59 & 0.56 \nl

M$_{d}$/L$_{R}$ (M$_{\sun}/$\L$_{\sun}$) & 
		2.59 & 15 & 0.28 & 2.36 & 0.21 \nl

\enddata

\end{deluxetable}

%% 
%% Table 2 
%%

\begin{deluxetable}{cccc}
\tablewidth{0pc}
\tablecaption{Parameters for the Dynamical Mass for the VII\,Zw\,466 Group}
\startdata\nl
\tableline \nl
Galaxy &  R$_{i}$\tablenotemark{a}~(kpc) &  $v_{zi}$\tablenotemark{b}
(km\,s$^{-1}$) &  q$_{i}$ = $\frac{R_{i}~v_{zi}^{2}}{G}$ (M$_{\sun}$)\nl
\tableline \nl
VII\,Zw\,466 & 33.7 & 325 &$8.00 \times 10^{11}$ \nl
G1 &   11.0  & -43 	& $4.57 \times 10^{9}$ \nl
G2 &  33.9  & 314	& $7.51 \times 10^{11}$ \nl
G3 & 188.4  & -97	& $3.98 \times 10^{11}$ \nl
G4 & 244.1  &  80	& $3.51 \times 10^{11}$ \nl
\tablenotetext{a}{Projected distance from the barycenter}
\tablenotetext{b}{Velocity offset from the systemic velocity of the
barycenter ($v_{cm}$=14,143 km\,s$^{-1}$).}
\enddata
\end{deluxetable}

\onecolumn

\begin{figure}[p]

\caption{Grey-scale image of VII\,Zw\,466 through a V--band filter taken
at Fick Observatory. The positions of the two newly discovered dwarfs
are marked as G3 and G4.}
\label{group}
\end{figure}

\begin{figure}[p]
\caption{Grey-scale image of VII\,Zw\,466 through a B--band filter
overlayed with a contour map of the integrated H\,I distribution. The
contour increment is 57.15~Jy\,beam$^{-1}$\,m\,s$^{-1}$ and the
level of the first contour is also 57.15 Jy\,beam$^{-1}$\,m\,s$^{-1}$.}
\label{f2}
\end{figure}

\begin{figure}[p]
\caption{The global H\,I profile of VII\,Zw\,466.}
\label{ring}
\end{figure}

\begin{figure}[p]
%\centering{\psfig{figure=7zwgraphs/g2.ps,width=6in}}
\caption{The global H\,I profile of the Companion Galaxy G2.}
\label{g2}
\end{figure}

\begin{figure}[p]
%\centering{\psfig{figure=7zwgraphs/g3.ps,width=6in}}
\caption{The global H\,I profile of the Dwarf Galaxy G3.}
\label{g3}
\end{figure}

\begin{figure}[p]
%\centering{\psfig{figure=7zwgraphs/g4.ps,width=6in}}
\caption{The global H\,I profile of the Dwarf Galaxy G4.}
\label{g4}
\end{figure}

\begin{figure}[p]
%\centering{\psfig{figure=7zwgraphs/fit.ps,width=6in}}
\caption{The H\,I velocity as a function of azimuth around the ring.
The solid circles were excluded from the fit.}

\label{7zwfit}
\end{figure}

\begin{figure}[p]
%\centering{\psfig{figure=7zwgraphs/dwarf2.ps,width=6in}}
\caption{Contour plots of the 27 channel maps of VII\,Zw\,466. The velocity of
each channel is displayed in the upper right corner. The contour
increment is 1.35$\times 10^{-4}$ Jy\,beam$^{-1}$ (0.5 $\sigma$ level)
and the lowest contour displayed is at 2.5$\sigma$. The solid crosses
indicate the position of the ring, G1, and G2, while the open cross
indicates the position of the background galaxy B1.}
\label{7zwchan}
\end{figure}

\begin{figure}
\caption{Two orthogonal views of a single timestep in a
numerical hydrodynamical model of the collision between two galaxies
with gas disks.  a) x-y plane, the initial midplane of the primary
disk; companion is seen edge-on at right.  b) y-z plane, the initial
plane of the companion, now seen at the bottom.  Both frames optimized
to illustrate the accretion streams.  Only 20\% of the 19,640 particles
in the primary disk, and 50\% of the 3240 particles in the companion
are plotted.  Line segments proportional to the projected direction
and magnitude of the velocity of selected particles, with the particle
plotted at the base, reveal the flow.  The selected particles include
all particles originating in the companion galaxy, with temperatures
of less than 10 times the initial temperature (i.e., $<$ 30,000--50,000K),
and located more than 0.5 (dimensionless) units from the z=0 plane.
See references in text for further details.}
\label{sph}
\end{figure}

\end{document}